\documentclass[aps,prl,showpacs,notitlepage,twocolumn,superscriptaddress,nofootinbib,preprintnumbers]{revtex4-2}
\usepackage{graphicx} 
\usepackage{epstopdf}
\usepackage{amsmath}
\usepackage{physics}
\usepackage{bm} 
\usepackage{amssymb}
\usepackage{xcolor}
\usepackage{array}
\usepackage{multirow}
\usepackage{lipsum}
\usepackage[T1]{fontenc}
\usepackage{tabu}
\usepackage{hhline}
\usepackage[english]{babel}
\usepackage{hyperref}

\renewcommand{\v}[1]{{\mathbf{\boldsymbol{#1}}}}

\renewcommand{\exp}{\text{exp}} 
\renewcommand{\det}{\text{det}}

\begin{document}

\title{Pairing-based graph neural network for simulating quantum materials} 
\author{Di~Luo}
\affiliation{Center for Theoretical Physics, Massachusetts Institute of Technology, Cambridge, MA 02139, USA}
\affiliation{The NSF AI Institute for Artificial Intelligence and Fundamental Interactions}
\affiliation{Department of Physics, Harvard University, Cambridge, MA 02138, USA}
\author{David~D.~Dai}
\affiliation{Department of Physics, Massachusetts Institute of Technology, Cambridge, MA 02139, USA}
\author{Liang~Fu}
\affiliation{Department of Physics, Massachusetts Institute of Technology, Cambridge, MA 02139, USA}

\date{\today}

\begin{abstract}
We develop a pairing-based graph neural network for simulating quantum many-body systems. Our architecture augments a BCS-type geminal 
wavefunction with a generalized pair amplitude parameterized by a graph neural network. Variational Monte Carlo with our neural network simultaneously provides an accurate,  flexible, and scalable method for simulating many-electron systems. We apply this method to two-dimensional semiconductor electron-hole bilayers and obtain accurate results on a variety of interaction-induced phases, including the exciton Bose-Einstein condensate, electron-hole superconductor, and bilayer Wigner crystal. Our study demonstrates the potential of physically-motivated neural network wavefunctions for quantum materials simulations. 
\end{abstract}

\maketitle

\textit{Introduction---} Many-electron systems offer a variety of intriguing emergent phenomena governed by the fundamental laws of quantum mechanics. Although Schrödinger’s equation already provides a “theory of everything” for condensed matter, actually solving Schrödinger’s equation is exponentially hard, rendering all but the simplest systems out of reach of direct solution. For almost a century, physicists have been attempting to breach this exponential wall by developing approximate numerical techniques such as Hartree-Fock and density functional theory. Although these techniques are powerful and provide important insights, they have their own limitations, especially in treating strongly correlated systems. Recent advances in artificial intelligence have opened up new directions for tackling quantum many-body physics~\cite{Carleo602}. Specifically, 
neural network wavefunctions can efficiently represent a large class of quantum states, including highly entangled ones~\cite{Deng_2017, gao2017efficient, Glasser_2018, Levine_2019, sharir2021neural, luo_inf_nnqs}. Promising applications of neural networks include computing ground state~\cite{chen2023autoregressive,robledo2022fermionic, chen2022simulating, doi:10.1126/science.aag2302, Hibat_Allah_2020, PhysRevLett.124.020503, Irikura_2020, PhysRevResearch.3.023095, Han_2020,ferminet,Choo_2019,rnn_wavefunction,paulinet,Glasser_2018,Stokes_2020,Nomura_2017,martyn2022variational,Luo_2019,PhysRevLett.127.276402, https://doi.org/10.48550/arxiv.2101.07243,luo2022gauge} and finite temperature properties, as well as simulating real-time dynamics~\cite{xie2021ab,wang2021spacetime,py2021, gutierrez2020real, Schmitt_2020,Vicentini_2019,PhysRevB.99.214306,PhysRevLett.122.250502,PhysRevLett.122.250501,luo_gauge_inv,luo_povm}.

In recent years, two-dimensional materials have emerged as an ideal platform for exploring quantum many-body phases in a controllable and tunable setting. For example, gate-tunable monolayer transition metal dichalcogenides (TMDs) realize the two-dimensional electron/hole gas, while recent experiments have created electron-hole bilayers using two TMD layers separated by an atomically thin tunneling barrier \cite{exciton1, exciton2}. These TMD bilayers feature separately tunable electron and hole densities, controllable barrier thickness, and a large exciton binding energy, promising a plethora of interaction-induced phases \cite{ZengBilayer, DaiTrion}. Very recently, transport and optical experiments reported the observation of perfect Coulomb drag in TMD bilayers with equal electron and hole densities, which demonstrates the dominance of exciton transport up to $20$ K~\cite{qi2023perfect,nguyen2023perfect}.    
Compared to the uniform electron gas, electron-hole bilayers exhibit a richer phase diagram due to the simultaneous presence of repulsion and attraction between charges, and are more difficult to study with conventional methods. Strong correlations render mean-field theory inadequate at low density, and more accurate methods sacrifice scalability. Additionally, as a continuum system the electron-hole bilayer cannot be easily treated with advanced numerical methods developed for lattice systems, such as exact diagonalization and density matrix renormalization group.

A powerful method for continuum systems simulation is variational Monte Carlo (VMC), which provides a variational bound for energy via trial wavefunction optimization. However, these trial wavefunctions are traditionally designed by hand for specific quantum phases and typically contain a small number of variational parameters, potentially introducing bias and limiting accuracy. In contrast, a general neural network wavefunction is capable of describing all possible phases of a system in a unified manner, thus eliminating bias and improving accuracy. The idea of a generic fermionic neural network wave function was initially introduced through neural network backflow~\cite{Luo_2019} on Slater determinant and Bogoliubov de Genne wavefunctions. It was further advanced by PauliNet~\cite{paulinet} and FermiNet~\cite{ferminet}, which developed the approach to quantum chemistry and demonstrated superior performance compared to other methods such as coupled cluster. Recently, further progress has been demonstrated on using fermionic neural network for simulations in continuum spaces~\cite{cassella2023discovering,luo2023artificial,pescia2023message,kim2023neural,lou2023neural,entwistle2023electronic,wilson2022wave,li2022ab,scherbela2022solving,adams2021variational}.

To study the phase diagram of electron-hole bilayers, it is essential to take into account of fermion pairing, which is responsible for the exciton superfluid. Generally speaking, number-projected BCS wavefunctions or equivalently Geminal wavefunctions, capture the physics of fermion pairing at a mean-field level. To further improve their accuracy especially in strong-coupling regime, an early attempt uses neural network as a Jastrow on top of a pair-product wavefunction~\cite{nomura2017restricted}.  
Later neural network backflow has been introduced transforming the BdG pairing wavefunction in Ref.~\cite{Luo_2019} and further development is made in the Geminal Network~\cite{xie2023deep} and the AGP FermiNet~\cite{lou2023neural}. Recently, a Pfaffian-Jastrow neural-network
quantum state with backflow transformation via message-passing network has been introduced to encode general pairing~\cite{kim2023neural}. We refer to Genimal neural network wavefunctions here in general as \textit{GemiNet}.

In this work, motivated by the physics insight, we develop a pairing-based neural wavefunction, which augments a real-space BCS mean-field ansatz with a momentum space generalized pair amplitude
parameterized by a graph neural network.  Our architecture is both accurate and is scalable to large systems through transfer learning. As a test, we apply the method to the electron-hole bilayer with balanced electron and hole densities, and we perform Hartree-Fock-Bogoliubov calculations for reference. We find that our Geminal neural wavefunction is quantitatively accurate across the whole parameter space, including the BEC, BCS, and Wigner crystal regimes. Our study paves the way for physically-motivated neural network wavefunctions to be developed and applied to other challenging continuum systems in quantum materials.

\begin{figure*}[t!]
  \centering
  \includegraphics[width=0.95\textwidth]{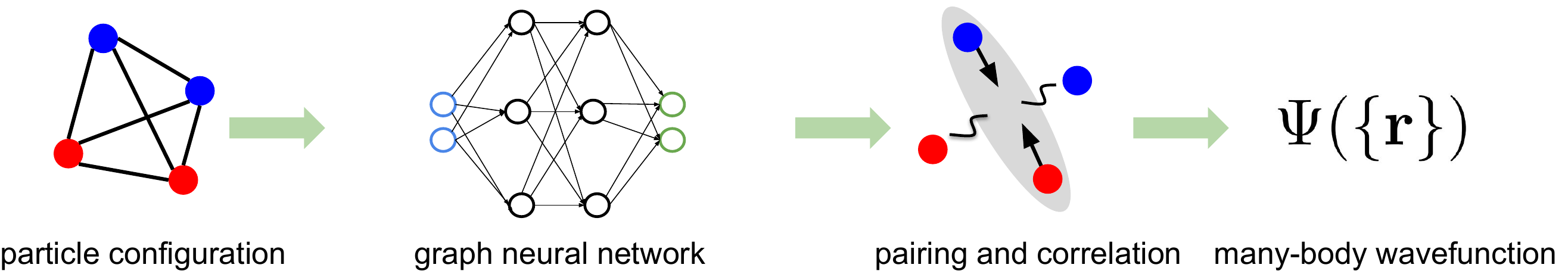}
  \caption{Schematic of our Geminal neural network wavefunction, a physically-motivated graph neural network that captures pairing and correlation in many-electron systems.}
  \label{fig:schematic}
\end{figure*}

\textit{Electron-hole bilayer---} We apply our GemiNet to study the semiconductor electron-hole bilayer with equal populations of electrons and holes \cite{theory1, theory3, theory2}. Working within the effective mass approximation and assuming $1/r$ Coulomb interactions, we have the Hamiltonian:
\begin{equation}\label{Hamiltonianh}
\begin{aligned}
    H =& -\frac{1}{2}\sum_{i}\bigg(\nabla^2_{ei} + \nabla^2_{hi}\bigg)  
    +\sum_{i<j}\frac{1}{|\v r^{e}_{i} - \v r^e_{j}|} \\ 
    &+ \sum_{i<j} \frac{1}{|\v r^{h}_{i} - \v r^{h}_{j}|} 
    -\sum_{i,j}\frac{1}{\sqrt{|\v r^{e}_{i} - \v r^{h}_{j}|^2 + d^2}},\\
\end{aligned}
\end{equation}
where $\v r^e_1, ..., \v r^e_N$ denotes the coordinates of electrons and $\v r^h_1, ..., \v r^h_N$ the coordinates of holes. For simplicity, we consider spinless electrons and holes with equal mass. All quantities are expressed in terms of their corresponding atomic units: lengths in terms of the Bohr radius $a_B = \frac{4\pi\epsilon\hbar^2}{m_e e^2}$ and energies in the Hartree $E_h = \frac{\hbar^2}{m_e a_B^2}$. Since our system is globally neutral, we work in the convention that there is no background charge in each layer. Compared to the inclusion of a uniform neutralizing background in each layer, our energies are higher by an energy per area of $2\pi dn^2$, where $d$ is the layer distance and $n=N/A$ is the exciton number density (see Supplementary Materials for details).

Our system is characterized by two length scales: the average interparticle distance $a=1/\sqrt{\pi n}$ and the layer distance $d$. Here, $a/a_B \equiv r_s$ characterizes the competition between kinetic and interaction energies, while $d/a_B$ determines the binding energy of a single exciton. For small $r_s$, the ground state is a two-component plasma dominated by the kinetic energy with weak-pairing described by the BCS theory. For large $r_s$ and small $d$, the ground state is a weakly interacting Bose-Einstein condensate of tightly-bound excitons. For large $r_s \gg 1$ and $d \gg a_B$, the ground state is two weakly coupled but aligned Wigner crystals on the two layers. The full phase diagram of the electron-hole bilayer as a function of $r_s$ and $d/a_B$ has been studied with various methods before.  

We first provide mean-field results for comparison with our GemiNet by performing self-consistent Hartree-Fock-Bogoliubov (HFB) calculations. The most general translationally-invariant mean-field ansatz for two fermion species of equal population is the Bogoliubov de Gennes (BdG) wavefunction:
\begin{equation}
\begin{aligned}
\Psi_{\text{BdG}}(\{\v r\})&= \det{[\psi(\v r^{e}_{i} - \v r^{h}_{j})]},\\
\end{aligned} \label{eq:BdG}
\end{equation}
where $\{\v r\}$ is the collection of all electron and hole coordinates, $\psi$ is an electron-hole pair function, and the determinant runs over the matrix with rows $i$ and columns $j$. For our calculations, we take $\psi$ to depend only on the electron-hole separation $\v r^e - \v r^h$, which captures translationally-invariant electron-hole pairing in states such as exciton superfluids but not translation symmetry breaking states such as crystals.

Following the standard treatment of pairing in Fermi systems, we perform all calculations in the grand canonical ensemble. For small layer distances $d$ for which the exciton is bound, the ground states at $n\rightarrow 0$ and $\infty$, respectively the free exciton gas and free electron-hole plasma, are both captured by the HFB ansatz. Therefore, we expect HFB to be accurate in both the high and low-density limits, while missing correlation effects for intermediate densities.

On the other hand, at large layer distance $d\gg a_B$ for which the exciton unbinds and low density $a \gg a_B$, the ground state is a Wigner crystal,  which is smoothly connected to the classical limit $a_B \rightarrow 0$. A natural mean-field description of the Wigner crystal is a product of an electron Slater determinant and a hole Slater determinant:
\begin{equation}\label{eq:det}
\begin{aligned}
\Psi_{\text{SD}}(\{\v r\}) =  \det{\big[\phi^e_{k} (\v r^e_{i})\big]} 
\det{\big[\phi^h_{l}(\v r^h_{j})\big]}, 
\end{aligned}
\end{equation}
where $\phi^{e}_k$ and $\phi^{h}_l$ denote a set of single-particle orbitals on the electron and hole layer respectively. The determinant runs over the matrix with rows $k$ or $l$ (orbital index) and columns $i$ or $j$ (particle index). To describe the Wigner crystal phase, these orbital wavefunctions $\phi^{e, h}$ must break translation symmetry. The use of mean-filed ansatzes of different forms---Eq.(\ref{eq:BdG}) and Eq.(\ref{eq:det})---makes unbiased comparison of competing phases difficult, and correlation is still missing.

\textit{Pairing-based Graph Neural Network with Physics Motivation---} As a starting point, we note that the BdG wavefunction can in principle also describe the Wigner crystal if the pair function $\psi(\v r^{e}_{i} - \v r^{h}_{j})$ is generalized to $\psi(\v r^{e}_{i}, \v r^{h}_{j})$, which depends on each coordinate individually instead of just their difference. This is because the identity $\det[M] \det[N] = \det [MN]$ allows the Wigner crystal wavefunction Eq. \ref{eq:det} to be expressed  as: 
\begin{equation}\label{eq:det combined}
\begin{aligned}
\Psi_{\text{SD}}(\{\v r\}) &=  \det{\big[\psi(\v r^{e}_{i}, \v r^{h}_{j})\big]},\\
\psi(\v r^{e}_{i}, \v r^{h}_{j}) &= \sum_{k}\phi^e_{k} (\v r^e_{ i})\phi^h_{k}(\v r^h_{ j}).\\
\end{aligned}
\end{equation}
While this generalized BdG ansatz can capture various phases in our system, it is cumbersome to optimize and is still inaccurate in intermediate regimes, necessitating the introduction of an ansatz that is both flexible and accurate.

To go beyond mean-field theory and incorporate strong correlations, we introduce a new variational wavefunction by replacing the original pair orbital $\psi(\v r^e_i, \v r^h_j)$ with a generalized pair orbital $\psi(\v r^e_i, \v r^h_j, \{\v r\})$:
\begin{align}
    \Psi_{\text{Geminal}}(\{\v r\}) &= \det[\psi(\v r^{e}_{i}, \v r^{h}_{j}; \{\v r_{/} \})] \\
    \psi(\v r^{e}_{i}, \v r^{h}_{j}; \{\v r_/\}) &= \sum_{\v k} \mathcal{G}_{\v k}(\v r^{e}_{i}, \v r^{h}_{j}; \{\v r_{/ } \})  e^{i \v k \cdot (\v r^e_i - \v r^h_j)} 
\end{align}
where $\mathcal{G}_{\v k} = g_{\v k} e^{i\v \theta_{\v k}}$ is generally complex-valued and can be regarded as the generalized pair amplitude in $\v k$ space. Here, $\{\v r\}$ denotes the configuration of all electrons and holes, and $\{\v r_/\}$ denotes the configuration of all electrons and holes except $\v r^e_i$ and $\v r^h_j$. To guarantee the antisymmetry of the many-body wavefunction, the generalized pair orbitals $\psi(\v r^e_i, \v r^h_j, \{\v r\})$ must be \textit{symmetric} under the exchange two electron or two hole coordinates, excluding $\v r^e_i$ and $\v r^h_j$. This condition is equivalent to $\mathcal{G}_{\v k}(\v r^{e}_{i}, \v r^{h}_{j}; \{\v r_{/ } \})$  being symmetric under the exchange of any two electrons or holes. 

Crucially, our wavefunction featuring the generalized pair amplitude $\mathcal{G}_{\v k}(\v r^{e}_{i}, \v r^{h}_{j}; \{\v r_{/ } \})$ can be regarded as a configuration dependent generalization of the BCS wavefunction $\sum_{\v k} g_{\v k}  e^{i \v k \cdot (\v r^e_i - \v r^h_j)}$, which our ansatz reduces to when $\mathcal{G}_{\v k}$ is configuration independent. Indeed, $\mathcal{G}_{\v k}(\v r^{e}_{i}, \v r^{h}_{j}; \{\v r_{/ } \})$ also includes the previous efforts on BCS wavefunction with backflow as special cases~\cite{maezono2013excitons}.
Furthermore, when the number of $\v k$ is equal to the number of particles $N$, and 
$\mathcal{G}_{\v k}$ is a factorizable function of the electron and hole configurations $\{\v r^e\}$ and $\{\v r^h\}$, our wavefunction reduces to a product of two determinants of generalized orbitals:
\begin{equation}
    \Psi(\{\v r\}) =  \det{\big[\phi_k^e(\v r^e_i;\{\v r_/ \})\big]} \det{\big[\phi_l^h(\v r^h_j;\{\v r_/\})\big]}.
\end{equation}
This structure also appears in FermiNet and PauliNet, and naturally describes a Wigner crystal as discussed previously. Therefore, our variational wavefunction captures both exciton superfluids and Wigner crystals in a unified manner and include beyond mean-field correlations. 

In this work, we utilize a graph neural network to generate the generalized pair amplitude $\mathcal{G}_{\v k}(\v r^{e}_{i}, \v r^{h}_{j}; \{\v r_{/ } \})$. Neural networks have been shown to be accurate and efficient approximators of high-dimensional functions. Recently, the message passing neural network have been introduced as a neural network backflow transformation for plane-wave orbital~\cite{pescia2023message} and the neural network parameterization in the Pfaffian pairing wavefunction~\cite{kim2023neural}. Inspired by the above advancement, we design a graph neural network with vertex attention for flexible parameterization and efficient optimization of $\mathcal{G}_{\v k}(\v r^{e}_{i}, \v r^{h}_{j}; \{\v r_{/ } \})$. Due to the graph neural network nature, transfer learning can be performed across systems of different particle numbers and parameters, so that results obtained from calculations on a small system can be used to pre-train calculations on larger systems. Motivated by physics consideration, our Geminal neural wavefunction targets the generalized pair amplitude in $\v k$ space instead of directly parameterizing the generalized pair orbital~\cite{lou2023neural}. This enables the reduction of parameters while maintains the generality. We also note that our construction is size extensive.

We construct $\mathcal{G}_{\v k}(\v r^{e}_{i}, \v r^{h}_{j}; \{\v r_{/ } \})$ in three stages. First, we input the particle configuration into a graph by defining a set of vertex vectors $\v v_\alpha^{0}$ and edge vectors $\v e_{\alpha\beta}^{0}$:
\begin{equation}
    \v v^{0}_{\alpha} = (v_0), \quad \v e_{\alpha\beta}^{0} = (\v r_{\alpha\beta}, ||\v r_{\alpha\beta}||, s_\alpha \cdot s_\beta),
\end{equation}
where each vertex corresponds to a particle and each edge corresponds to a pair of particles. $\alpha$ and $\beta$ are particle (electron and hole) indices running from $1$ to $2N$, $v_0$ is a randomly initialized parameters, $||\v r_{\alpha\beta}|| = ||\v r_{\alpha} - \v r_{\beta}||$ is the distance between particles $\alpha$ and $\beta$, and $s_\alpha=\pm 1$ for electrons and holes respectively. To satisfy the periodic boundary condition, we use the $\sin$ and $\cos$ function embedding (see Supplementary Materials). Our input structure preserves both our system's translation and permutation symmetry. We further construct $\v v^{1}_{\alpha} = [\v v^{0}_{\alpha}, v_A]$ and $\v e_{\alpha\beta}^{1} = [\v e_{\alpha\beta}^{0},v_B]$ with $v_A$ and $v_B$ being randomly initialized parameters and $[\cdots]$ is the concatenation operation. 

\begin{figure}[t!]
  \centering
  \includegraphics[width=0.45\textwidth]{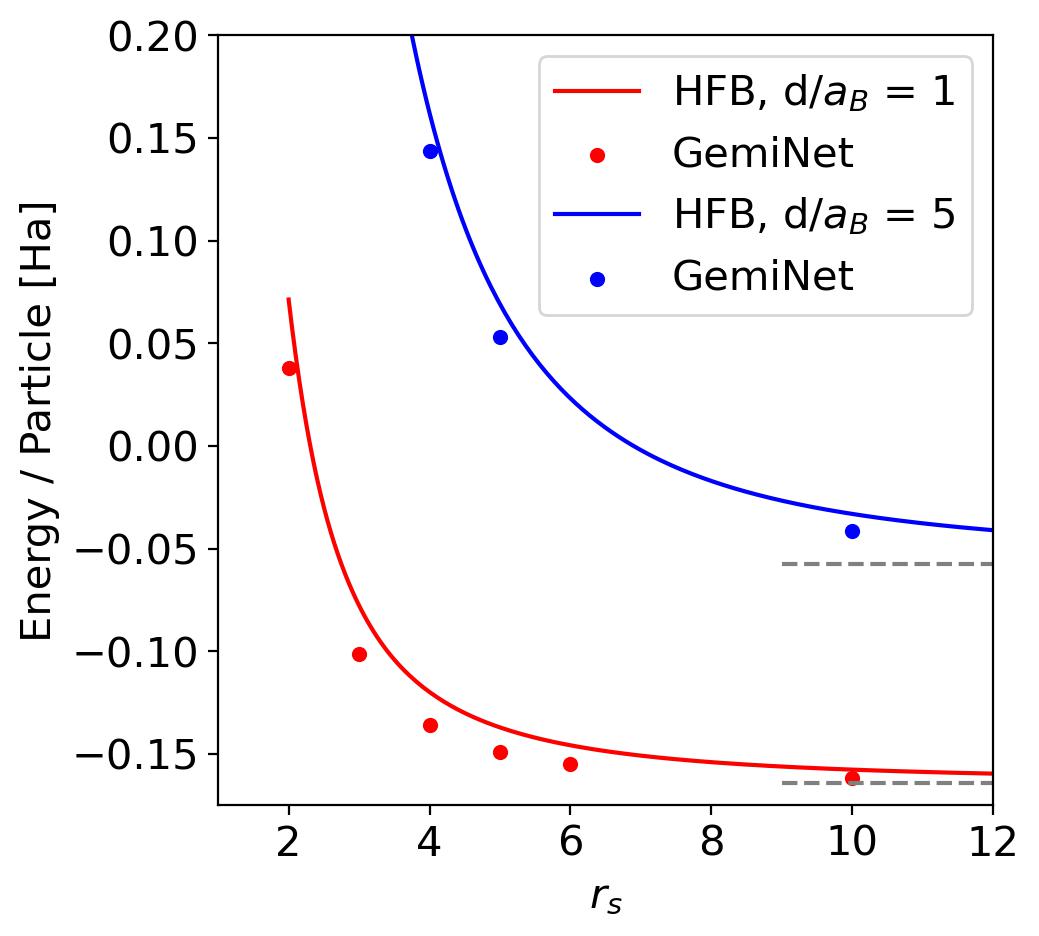}
  \caption{Comparison of mean-field and neural network energies for two values of $d$ and a range of $r_s$. The neural network and mean-field results have close agreement for high and low densities, but are most different for intermediate parameters when correlation is strongest. The dashed grey lines mark half the single-exciton binding energy for reference.}
  \label{fig:energy comparison}
\end{figure} 

\begin{figure*}[t!]
  \centering
  \includegraphics[width=0.85\textwidth]{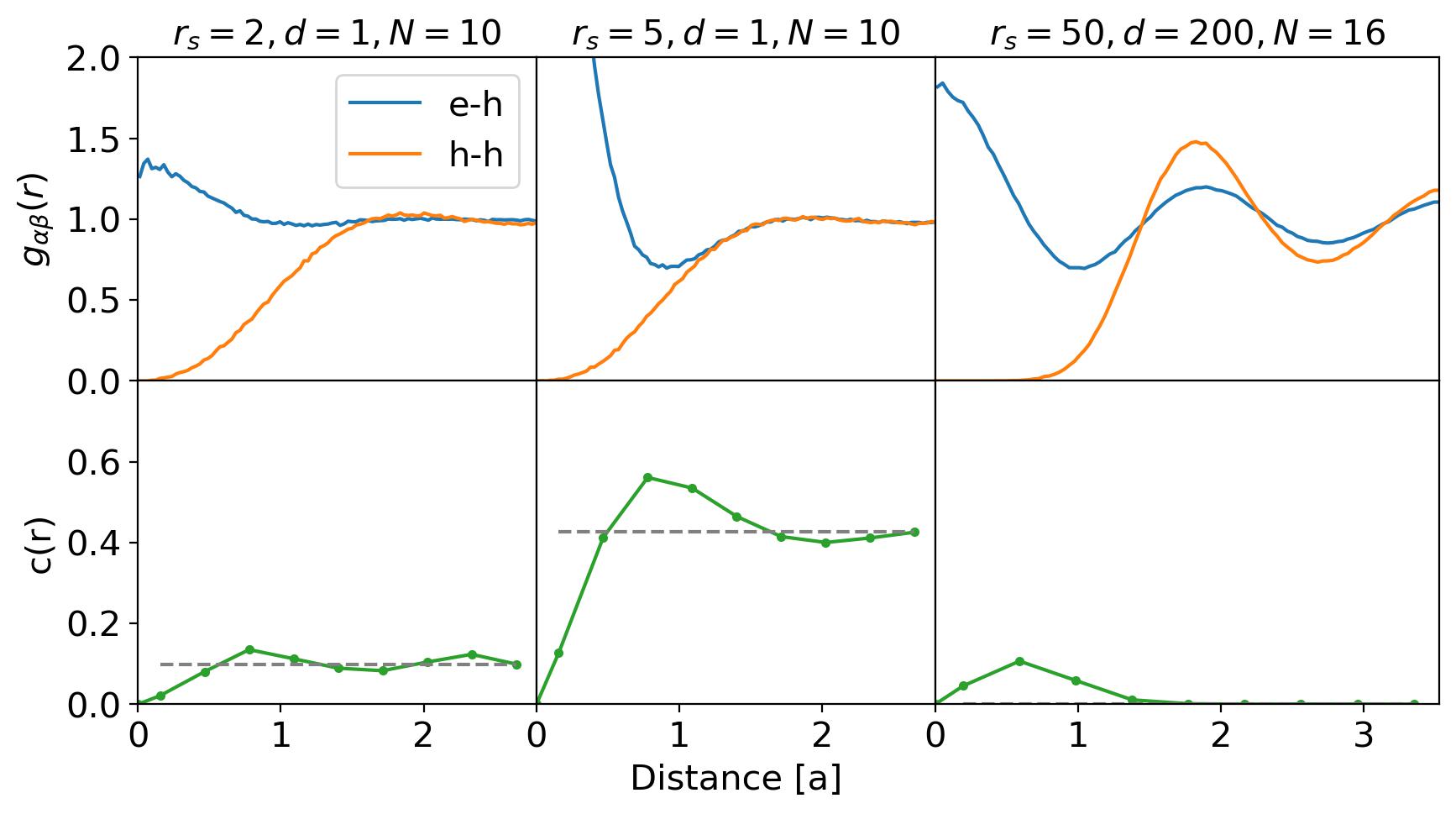}
  \caption{Pair correlation functions and condensate functions for three distinct regimes. The systems contain $N$ electrons and $N$ holes. $e$-$h$ and $h$-$h$ are the electron-hole and hole-hole pair correlations. The condensate \textit{fraction} is the $r\rightarrow\infty$ limit of the condensate function.}
  \label{fig:three panel}
\end{figure*} 

Next, we embed the vertex and edge vectors into a high-dimensional latent space and process them iteratively. The $l$-th iteration ($l \geq 2$) is:
\begin{align}
    \v m_\alpha^Q &= \sum_\beta W_Q^l \v e_{\alpha\beta}^{l-1}, \quad \v m_\beta^K = \sum_\alpha W_K^l \v e_{\alpha\beta}^{l-1},\\
    \label{Eq:vertex}
    \v m_{\alpha\beta}^{l} &= F^{l}(\v m^Q_\alpha \otimes \v m^K_\beta) \odot G^{l}(\v e_{\alpha\beta}^{l-1}),\\
    \v e_{\alpha\beta}^{l} &= [H^{l}([\v m_{\alpha\beta}^l, \v v_{\alpha\beta}^{l-1}]), \v e_{\alpha\beta}^{0}],\\
    \v v_{\alpha}^{l} &= [U^{l}([\sum_j \v e_{\alpha\beta}^{l},\v v_{\alpha}^{l-1}]),\v v_{\alpha}^{0}],
\end{align}
where $F^{l},G^{l},H^{l}, U^{l}$ are two-layer fully-connected neural networks with GELU activation function and hidden dimension 32, and $W^l_Q$ and $W^l_K$ are embedding matrices. Here, $\otimes$ is the tensor product, $\odot$ is the element-wise multiplication. The above construction guarantees the permutation equivalence for GNN during the update. We note that our GNN is inspired by the recent development of message passing neural network quantum state~\cite{pescia2023message,kim2023neural}. Meanwhile, we design a new vertex attention operation in Eq. \ref{Eq:vertex} to capture correlations of all particles with a low $O(N^2)$ complexity.

Finally, after $T$ iterations, we project the graph's high-dimensional latent space features into lower-dimension space to construct $\mathcal{G}_{\v k}$,
\begin{align}
    \v p_\alpha &= C v_{\alpha}^L, \quad \v f_{ij} = [\v p^e_i,\v p^h_j, \v p^e_i \odot \v p^h_j],\\
    \mathcal{G}_{\v k}(\v r^{e}_{i}, \v r^{h}_{j}; \{\v r_{/ } \}) &=  \exp(-\v k \cdot W(\v f_{ij}) + R([v^L_{e_i},v^{L}_{h_j},k]))
\end{align}
where $C$ is a randomly initialized embedding matrix. $R$ is a two-layer fully-connected neural networks with GELU activation function, hidden dimension 32 and output dimension 1. $W$ is a complex-valued linear projection matrices with output dimension 2. The details can be referred to the Supplementary Materials.

We minimize the variational energy using Monte Carlo sampling and the natural gradient method~\cite{sorella1998green}. Evaluating the determinant has complexity $O(N^3)$, and optimization incurs an addition factor of $O(N)$ due to the time needed for the Markov chain to equilibrate, yielding a formal complexity of $O(N^4)$.

\textit{Results---} After optimizing the wavefunction, we compute physical observables such as energy and various correlation functions. Here, we evaluate the energy, electron-electron and electron-hole correlation functions, and the condensate fraction.

Fig.\ref{fig:energy comparison} shows the ground state energy per particle as a function of $r_s$ for two layer distances $d=1a_B$ and $5a_B$, obtained from both our GemiNet and Hartree-Fock-Bogoliubov mean-field calculation. In all cases, our GemiNet yields lower energies, and the improvement is particularly significant at intermediate $r_s$ where strong correlation effect is expected.       

The translation-rotation average of the pair correlation function between species $\alpha$ and $\beta$ is given as :
\begin{equation}\label{observable definition}
    g_{\alpha \beta}(r) = \frac{\int |\Psi(\v R)|^2 \delta^{(2)}(\v r^{\alpha}_1 - \v r^{\beta}_2 - \v r') \delta(|\v r'|-r) \text{d}\v r' \text{d}\v R}{L^2 \cdot 2 \pi r \int |\Psi(\v R)|^2 \text{d}\v R},
\end{equation}
and the condensate function can be expressed in terms of the one-body and two-body density matrices:
\begin{equation}
    \rho^{(1)}_{\alpha}(r) = \frac{N \int |\Psi(\v R)|^2 \frac{\Psi(\v r^{\alpha}_1+\v r')}{\Psi(\v R)} \delta(|\v r'|-r) \text{d}\v r' \text{d}\v R}{L^2 \cdot 2 \pi r \int |\Psi(\v R)|^2 \text{d}\v R},
\end{equation}
\begin{equation}
    \rho^{(2)}(r) = \frac{ N^2\int |\Psi(\v R)|^2 \frac{\Psi(\v r^e_1 + \v r',\v r^h_1 + \v r')}{\Psi(\v R)} \delta(|\v r'|-r) \text{d}\v r' \text{d}\v R}{L^2 \cdot 2 \pi r \int |\Psi(\v R)|^2 \text{d}\v R},
\end{equation}
\begin{equation}
    c(r) = \frac{L^2}{N} (\rho^{(2)}(r)-\rho^{(1)}_{e}(r)\rho^{(1)}_{h}(r)).
\end{equation}
Here, $\text{d}\v R$ denotes integration over the configuration of all electrons and holes, $\Psi(\v R)$ denotes the wavefunction evaluated at the configuration $\v R$, and $\Psi(\v r^{\alpha}_1+\v r')$ is shorthand for the wavefunction evaluated at $\v R$, except that coordinate $\v r^\alpha_1$ is replaced with $\v r^\alpha_1 + \v r'$, and likewise for $\Psi(\v r^e_1 + \v r',\v r^h_1 + \v r')$. The condensate fraction is defined as $c = \textup{lim}_{r \rightarrow \infty} c(r)$. 

In Fig. \ref{fig:three panel}, we present results for three distinct parameter regimes in our system: the weakly-paired electron-hole liquid, the exciton Bose-Einstein Condensate, and Wigner crystal. The left panel shows the observables for the BCS regime, where $r_s = 2$ and $d=1a_B$. Here, the pairing is weak because the exciton radius is larger than the average inter-exciton distance $a$, which is consistent with the relatively small electron-hole correlation peak at short distance, as well as the small condensate fraction $\sim 10\%$. 

The middle panel shows the correlation functions and condensate function for the BEC regime, where $r_s = 5$ and $d=1 a_B$. Here, the electron-hole correlation at short range is very strong, consistent with the exciton radius being smaller than the average inter-exciton distance. Also, a large condensate fraction $> 40\%$ is found. In both BEC and BCS regimes, the electron-electron correlation goes from zero at short distance (as expected from repulsion) to $1$ at long-distance without strong oscillations, indicating the lack of crystallization. 

The right panel shows the observables for the expected Wigner crystal regime at $r_s = 50$ and a large layer distance $d = 200a_B$. In both the electron-hole and electron-electron correlations, two prominent peaks are observed at inter-particle distances $1.9 a$ and $3.5 a$  at the largest system size in our study with 32 particles. The presence of second peak supports our interpretation of this state as a bilayer Wigner crystal with long-range order in the thermodynamic limit. For comparison, strongly correlated liquids such as ${}^4$He also have correlation peaks at short-distance, but not at larger distances. The large short-range electron-hole correlation is due to the alignment of the Wigner crystals in the two layers and is not a sign of superfluidity, which is consistent with the negligible condensate fraction.

Our analysis of the correlation functions demonstrates the capability of using one unified neural network ansartz to discover distinct phases of a quantum many-body system with remarkably high accuracy, in contrast to the traditional method of using separate ansatzes for each phase and comparing which ansatz has the lowest energy.

\textit{Conclusion---} In this work, we develop a pairing-based graph neural network for simulating quantum phases of matter. We construct a Geminal neural wavefunction using physics intuition by starting with a Bogoliubov de Gennes mean-field wavefunction and augmenting it with a generalized pair amplitude by a graph neural network. Our approach is flexible, quantitatively accurate, and scalable. We test the method on the semiconductor electron-hole bilayer and find that it accurately captures various phases, including the Bose-Einstein condensate, electron-hole superconductor, and bilayer Wigner crystal, yielding both the expected correlation functions and ground state energies significantly lower than the mean-field results. Our study encourages further development of physically-motivated neural network wavefunctions and their application to challenging continuum systems in quantum materials.

\textit{Acknowledgement---} We thank helpful discussions with Gabriel Pescia, Wan Tong Lou, Jane Kim, Kyung-Su Kim, Ryan Levy, Bryan Clark, Michael Scherbela, Leon Gerard, M. T. Entwistle, Aidan P. Reddy, Zhuo Chen, Yin Lin, Nicholas Gao, Yuanqi Du and Giuseppe Carleo.
We acknowledge the MIT SuperCloud for providing the computing resources used in this paper. 
LF acknowledges the support from National Science Foundation (NSF) Convergence Accelerator Award No. 2235945.
DL acknowledges support from the NSF AI Institute for Artificial Intelligence and Fundamental Interactions (IAIFI) and the U.S. Department of Energy, Office of Science, National Quantum Information Science Research Centers, Co-design Center for Quantum Advantage (C2QA) under contract number DE-SC0012704. DDD was supported by the Undergraduate Research Opportunities Program at MIT.

\bibliography{bibliography,references_ml}


\appendix

\clearpage

\onecolumngrid
\begin{center}
	\noindent\textbf{Supplementary Material}
	\bigskip
		
	\noindent\textbf{\large{}}
\end{center}

\twocolumngrid

\section{Exciton Binding Energy}

To verify that the neural-network behaves correctly in the dilute limit, we solve the two-body electron-hole problem on a rectangular grid (i.e. derivatives replaced by finite differences) to calculate the binding energy of a single exciton at $d=1a_B$ and $d=5a_B$. The convergence metric is the minimum value of the wavefunction (on the border of the box) divided by the maximum value of the wavefunction (in the box's center). Our results are summarized and compared to our GemiNet's results at $r_s = 50$, which is well in the dilute limit, in the table below.
\begin{table}[h!]
\centering
\begin{tabular}{|c|c|c|}
 \hline
 Quantity & Grid $d = 1$&Grid $d = 5$\\ \hline
 Grid Size & $1000 \times 1000$ & $1000 \times 1000$\\ \hline
 Box Size &   $50a_B \times 50 a_B$ & $100a_B \times 100 a_B$\\ \hline
 Resolution & $0.05 a_B$ & $0.1 a_B$ \\ \hline
 Convergence & $<0.00001$  & $<0.00001$ \\  \hline
 Energy & $-0.3277 E_h$ & $-0.1144 E_h$ \\ \hline
 Exciton Radius & $2.76 a_B$ & $6.05 a_B$ \\ \hline
\end{tabular}
\label{tab:grid}
\end{table}
The energy per exciton at $r_s = 50$ and $d = 1$ calculated by our GemiNet is $-0.3260$, slightly above the single-exciton binding energy $-0.3277$ as expected due to repulsion. The energy per exciton at $r_s = 50$ and $d=5$ calculated by our GemiNet is $-0.1138$, slightly above the single-exciton binding energy $-0.1144$ as expected due to repulsion. These calculations indicate that our GemiNet functions as expected in the thermodynamic limit.

\section{Hartree-Fock Bogoliubov Calculations}

To provide a mean-field control for the neural-network results, we perform Hartree-Fock-Bogoliubov (HFB) calculations. The electrostatic attraction between electrons and holes makes the mean-field theory of the bilayer at $1:1$ density ratio similar to an $s$-wave superconductor with the electron's spin replaced by an electron/hole layer pseudospin. The most general translationally-invariant mean-field ansatz for two fermion species of equal population is the Bogoliubov de Gennes (BdG) wavefunction:
\begin{equation}\label{particle conserving ansatz}
\Psi(\{\v r\}) = \det{\bigg[\psi(\v r^e_i - \v r^h_j)\bigg]},
\end{equation}
where $\v r^e_i$ is the position of the $i$th electron, $\v r^h_j$ is the position of the $j$th hole, and $\psi(\v r)$ is a to-be-determined pairing wavefunction. Our ansatz allows pairing between electrons and holes but does not break translation symmetry, as expected of an exciton superfluid. 

In second quantization, the BdG ansatz becomes:
\begin{equation}
\begin{split}
\psi(\v r) &= \sum_{\v k} g(\v k) e^{i \v k \cdot \v r},\\
\Omega &= \sum_{\v k} g(\v k) a_{\v k}^\dagger b_{-\v k}^\dagger,\\
|\Psi\rangle &= \Omega^N|\text{vac}\rangle,\\
\end{split}
\end{equation}
where $a_{\v k}^\dagger$ is the momentum $\v k$ electron creation operator, and $b_{\v k}^\dagger$  is the momentum $\v k$ hole creation operator. As in the case for superconductivity, to make calculating the variational energy tractable we work in the grand canonical ensemble and control the particle number indirectly with the chemical potential $\mu$ (relaltive number fluctuations $\propto 1\sqrt{N}$ are suppressed in the thermodynamic limit). Up to normalization, our ansatz becomes:
\begin{equation}\label{particle nonconserving}
    \begin{split}
        \Psi &\propto \exp{(\Omega)}|\text{vac}\rangle,\\
        &= \prod_{k} \exp\bigg(g(\v k) a_{\v k}^\dagger b_{- 
            \v k}^\dagger\bigg) |\text{vac}\rangle,\\
        &= \prod_{k} \bigg(1 + g(\v k) a_{\v k}^\dagger b_{- 
            \v k}^\dagger\bigg) |\text{vac}\rangle.
    \end{split}
\end{equation}

Because our system is time-reversal symmetric, $g(\v k)$ can be taken to taken to be real. We can parameterize the normalized BdG ansatz using angles $\theta_{\v k}$:
\begin{equation}\label{normalized ansatz}
    \Psi = \prod_{k} \bigg(\cos{(\theta_{\v k} / 2)} + \sin{(\theta_{\v k} / 2)}  
        a_{\v k}^\dagger b_{-\v k}^\dagger\bigg) |\text{vac}\rangle.
\end{equation}
Letting the system area be $A$ and the interlayer spacing be $d$, the second-quantized electron-hole Hamiltonian is:
\begin{equation}\label{BCS Hamiltonian}
    \begin{split}
    H - \mu N=& \sum_{\v k} (\epsilon_{\v k} - \mu) (a_{\v k}^\dagger a_{\v k} + b_{\v k}^\dagger b_{\v k})\\
    &+ \frac{1}{2A}\sum_{\v k \v k' \v q} v_\text{intra}(\v q) a_{\v k+q}^\dagger a_{\v k'-q}^\dagger a_{\v k'} a_{\v k}\\
    &+ \frac{1}{2A}\sum_{\v k \v k' \v q} v_\text{intra}(\v q) b_{\v k+q}^\dagger b_{\v k'-q}^\dagger b_{\v k'} k_{\v k}\\
    &- \frac{1}{A}\sum_{\v k \v k' \v q} v_\text{inter}(\v q) a_{\v k+q}^\dagger b_{\v k'-q}^\dagger b_{\v k'} a_{\v k},\\
    \end{split}
\end{equation}
where $N$ is the total particle number, $\epsilon_{\v k} = k^2/2$ is the dispersion, $v_\text{intra}(\v q) = 2\pi / q$ is the Fourier transform of the intralayer interaction,  and $v_\text{inter}(\v q) = 2\pi \exp{(-qd)} / q$ is the Fourier transform of the interlayer interaction. It is straightforward to show that the average occupancies are $n_{\v k} = \langle a_{\v k}^\dagger a_{\v k} \rangle = \langle k_{\v k}^\dagger k_{\v k} \rangle = \sin^2(\theta_{\v k}/2)$ and the anomalous averages are $\langle b_{-\v k}a_{\v k} \rangle = \langle a_{\v k}^\dagger b_{-\v k}^\dagger \rangle = \sin(\theta_{\v k} / 2) \cos(\theta_{\v k} / 2) = \sin(\theta_{\v k}) / 2$. 

Then the variational HFB energy is:
\begin{equation}\label{BCS energy}
    \begin{split}
        F =& 2 \sum_{\v k} (\epsilon_{\v k} - \mu) \sin^2{(\theta_{\v k} / 2)}\\
        &- \frac{1}{A}\sum_{\v k \v k'}v_\text{intra}(\v k-\v k')\sin^2(\theta_{\v k}/2) \sin^2(\theta_{\v k}'/2)\\
        &- \frac{1}{A}\sum_{\v k \v k'}v_\text{inter}(\v k-\v k')\frac{\sin(\theta_{\v k})\sin(\theta_{\v k}')}{4},
    \end{split}
\end{equation}
which can be easily optimized with respect to the angles $\theta_{\v k}$ using standard techniques such as conjugate gradient. This yields an upper bound to the bilayer's energy, as well as other physical observables such as the plane-wave occupancies and quasiparticle spectrum.

As is generally true for mean-field theories, we explicitly show here that the HFB ansatz is exact in both the high and low-density limits. In the high-density limit, where the kinetic energy dominates both interlayer and intralayer interactions, the ground state reduces to a product of plane-wave Slater determinants, which can be captured by the HFB ansatz:
\begin{equation}\label{eq:high density reduce}
\begin{aligned}
\Psi_{\text{Fermi Sea}}(\{\v r\}) 
&= \det{\big[e^{i\v k_a \cdot \v r^e_i}\big]} \det{\big[e^{i\v k_b \cdot \v r^h_j}\big]},\\\
&= \det{\big[e^{i\v k_a \cdot \v r^e_i}\big]} \det{\big[e^{-i\v k_b \cdot \v r^h_j}\big]},\\
&= \det{\big[\sum_{k < k_F} e^{i\v k (\v r^e_i - \v r^h_j)}\big]},\\
\end{aligned}
\end{equation}
where $\sum_{k < k_F} \exp{(i\v k (\v r^e_i - \v r^h_j))}$ is identified with the pairing function $\psi$ in Eq. \ref{particle conserving ansatz}. The equality between the first and second line follows because $\v k$ is inside the Fermi sea if and only if $- \v k$ is. The Fermi sea ansatz corresponds to $\theta_{\v k} = \pi$ if $k < k_F$ and $\theta_{\v k} = 0$ if $k > k_F$, and the variational energy is the sum of the Fermi sea's kinetic energy and the Hartree-Fock exchange energy (equivalent to first-order perturbation theory in the interaction).

Below the $d$ at which the exciton unbinds, at at low-density when the average inter-particle distance is much larger than the exciton radius, we expect the ground state to be a non-interacting condensate of excitons. This corresponds to identifying $\psi$ in Eq. \ref{particle conserving ansatz} with the wavefunction of an isolated electron-hole pair. Although not immediately obvious, this can also be recovered directly from the variational HFB energy in Eq. \ref{BCS energy}.

The occupancy of wavevector $\v k$ is $n_{\v k} = \sin^2(\theta_{\v k}/2)$, which may be approximated as $\theta_{\v k}/2$ in the dilute limit since all occupancies are small. Then the variational HFB energy reduces to:
\begin{equation}\label{approximated HFB energy}
    \begin{split}
F =& 2 \sum_{\v k} (\epsilon_{\v k} - \mu) \bigg(\frac{\theta_{\v k}}{2}\bigg)^2\\
&- \frac{1}{A}\sum_{\v k \v k'}v_\text{inter}(\v k-\v k')\bigg(\frac{\theta_{\v k}}{2}\bigg)\bigg(\frac{\theta_{\v k'}}{2}\bigg) + O(\theta^4). \end{split}
\end{equation}
The now quadratic energy can be minimized analytically to yield:
\begin{equation}\label{exciton integral equation}
2k^2 \theta_{\v k} - \frac{1}{A}\sum_{\v k'}v_\text{inter}(\v k-\v k')\theta_{\v k'} = (2 \mu) \theta_{\v k},
\end{equation}
which is equivalent to Schr\"odinger;'s equation for an electron-hole pair in momentum space. The exciton chemical potential $2\mu$ is identified with the energy of a single exciton. Thus, we see that HFB reduces to the two-body problem in the dilute limit.

\section{Capacitance Energy}

Because both the intralayer and intralayer interactions are long ranged, their Fourier transforms are divergent at $\v q = \v 0$. Ordinarily, this is corrected by simply setting $v(\v q=\textbf{0}) = 0$. This can be interpreted as the effect of a uniform background charge, which is necessary in most situations to render the system globally neutral and make the thermodynamic limit reasonable. 

However, in our case the number of electrons and holes is equal, and on physical grounds we expect the system's thermodynamic limit to be sensible without needing a background charge. In this case, the divergence in the intralayer and interlayer potential cancel each other. The contribution of the interaction's $\v q = \v0$ Fourier component is:
\begin{equation}\label{BCS Hamiltonian q0}
    \begin{split}
    H(\v q = \v 0)= &+\frac{1}{2A}\sum_{\v k \v k'} v_\text{intra}(\v q = \v 0) a_{\v k}^\dagger a_{\v k'}^\dagger a_{\v k'} a_{\v k}\\
    &+ \frac{1}{2A}\sum_{\v k \v k'} v_\text{intra}(\v q = \v 0) b_{\v k}^\dagger b_{\v k'}^\dagger b_{\v k'} k_{\v k}\\
    &- \frac{1}{A}\sum_{\v k \v k'} v_\text{inter}(\v q = \v 0) a_{\v k}^\dagger b_{\v k'}^\dagger b_{\v k'} a_{\v k}.\\
    \end{split}
\end{equation}
Using the canonical anticommutators, we may rearrange the quartic term $a_{\v k}^\dagger a_{\v k'}^\dagger a_{\v k'} a_{\v k}$ into $(a_{\v k}^\dagger a_{\v k}) (a_{\v k'}^\dagger a_{\v k'}) - \delta_{\v k, \v k'} a_{\v k}^\dagger a_{\v k}$.

The second term is a subleading self-energy correction that may be ignored in the thermodynamic limit. The $\v q = \v 0$ part of the Hamiltonian becomes:
\begin{equation}\label{BCS Hamiltonian q0 thermo}
    \begin{split}
    H(\v q = \v 0)= &+\frac{1}{2A} v_\text{intra}(\v q = \v 0) \bigg(\sum_{\v k} a_{\v k}^\dagger a_{\v k}\bigg) \bigg(\sum_{k'} a_{\v k'}^\dagger a_{\v k'}\bigg)\\
    &+\frac{1}{2A} v_\text{intra}(\v q = \v 0) \bigg(\sum_{\v k} b_{\v k}^\dagger b_{\v k}\bigg) \bigg(\sum_{k'} b_{\v k'}^\dagger b_{\v k'}\bigg)\\
    &-\frac{1}{A}v_\text{inter}(\v q = \v 0) \bigg(\sum_{\v k} a_{\v k}^\dagger a_{\v k}\bigg) \bigg(\sum_{k'} b_{\v k'}^\dagger b_{\v k'}\bigg).\\
    \end{split}
\end{equation}
As the Hamiltonian conserves both the total electron number $\sum_{\v k} a_{\v k}^\dagger a_{\v k}$ and total hole number $\sum_{\v k} b_{\v k}^\dagger b_{\v k}$, and we work in the sector where both the electron and hole populations are equal to $N$, we may replace total number operators with their eigenvalues to write:
\begin{equation}\label{capacitance total}
    H(\v q = \v 0)= [v_\text{intra}(\v q = \v 0) - v_\text{inter}(\v q = \v 0)] N^2.
\end{equation}
This is diagonal and makes a contribution to the energy density $u(n)$, where $n=N/A$ is the exciton density:
\begin{equation}\label{capacitance density}
    u_{\v q = \v 0}(n)= n^2 \lim_{\v q \rightarrow \v 0} \frac{2\pi}{q}(1 - \exp{(-qd)} = 2 \pi d n^2.
\end{equation}
Since each unit of area has $2n$ particles, the capacitance energy per particle is $\pi d n$. For the correct interpretation of the bilayer's total energy density, especially when comparing energies of different densities, it is crucial to correctly and carefully treat the $\v q = \v 0$ capacitance energy. If the above expression (given in Hartree atomic units) is converged to SI units, we recover the classical capacitance energy between two parallel plates.

\section{Finite-Size Extrapolation}\

Because we perform HFB calculations in the grand canonical ensemble, we need to extrapolate to the thermodynamic limit to remove errors caused by relative number fluctuations. We performed calculations in large square boxes of sizes $L/a_B=300,500,700$ with the plane-wave cutoff chosen adaptively to ensure that the ratio between the the occupation of the minimally and maximally occupied plane-wave states was below $10^{-5}$. In general, we did not observe the plane-wave cutoff to have a large effect of the energy per exciton. In contrast, the energy per exciton changed significantly with system size.

All else held constant, we found a robust linear trend between the energy per exciton and relative number fluctuation $\sigma(N)/\langle N\rangle$, where $\sigma(N)$ and $\langle N \rangle$ respectively are the standard deviation and average of the particle number. The data reported in the main text come from fitting a line to the relationship between the energy per exciton and relative fluctuation and taking the value of the fit at zero fluctuation. Because our HFB calculations were at discrete number of points, and the precise value of $r_s$ slightly changed with different $L$ since we control $\mu$ not $r_s$, we first fit the HFB calculation results at a given $L$ to a cubic spline to interpolate to values of $r_s$ not explicitly sampled. All energies reported in this section have no capacitance energy because our calculations were performed in momentum-space where the $\v q= \v 0$ interaction must be explicitly removed to remove infinite matrix elements. 
 
\begin{figure}[t]
  \centering
  \includegraphics[width=0.50\textwidth]{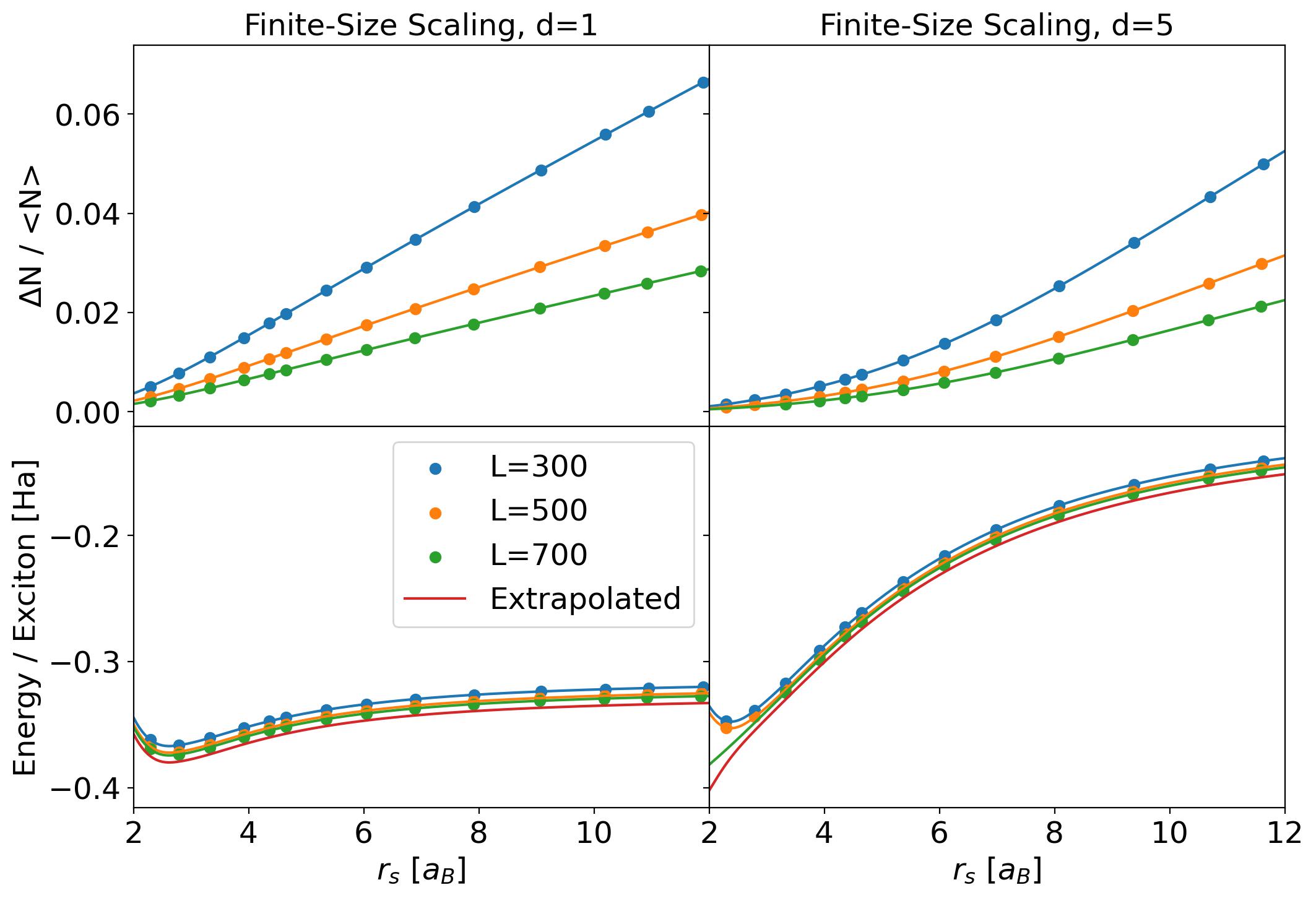}
  \caption{Finite-size analysis for Hartree-Fock Bogoliubov calculations for $d=1$, $d=5$, and a range of $r_s$. The dots are distinct HFB calculations, and the smooth lines connecting them is a cubic spline interpolation. All energies here have the capacitance energy removed.}
  \label{fig:extrapolation d5}
\end{figure} 
\begin{figure}[t]
  \centering
  \includegraphics[width=0.45\textwidth]{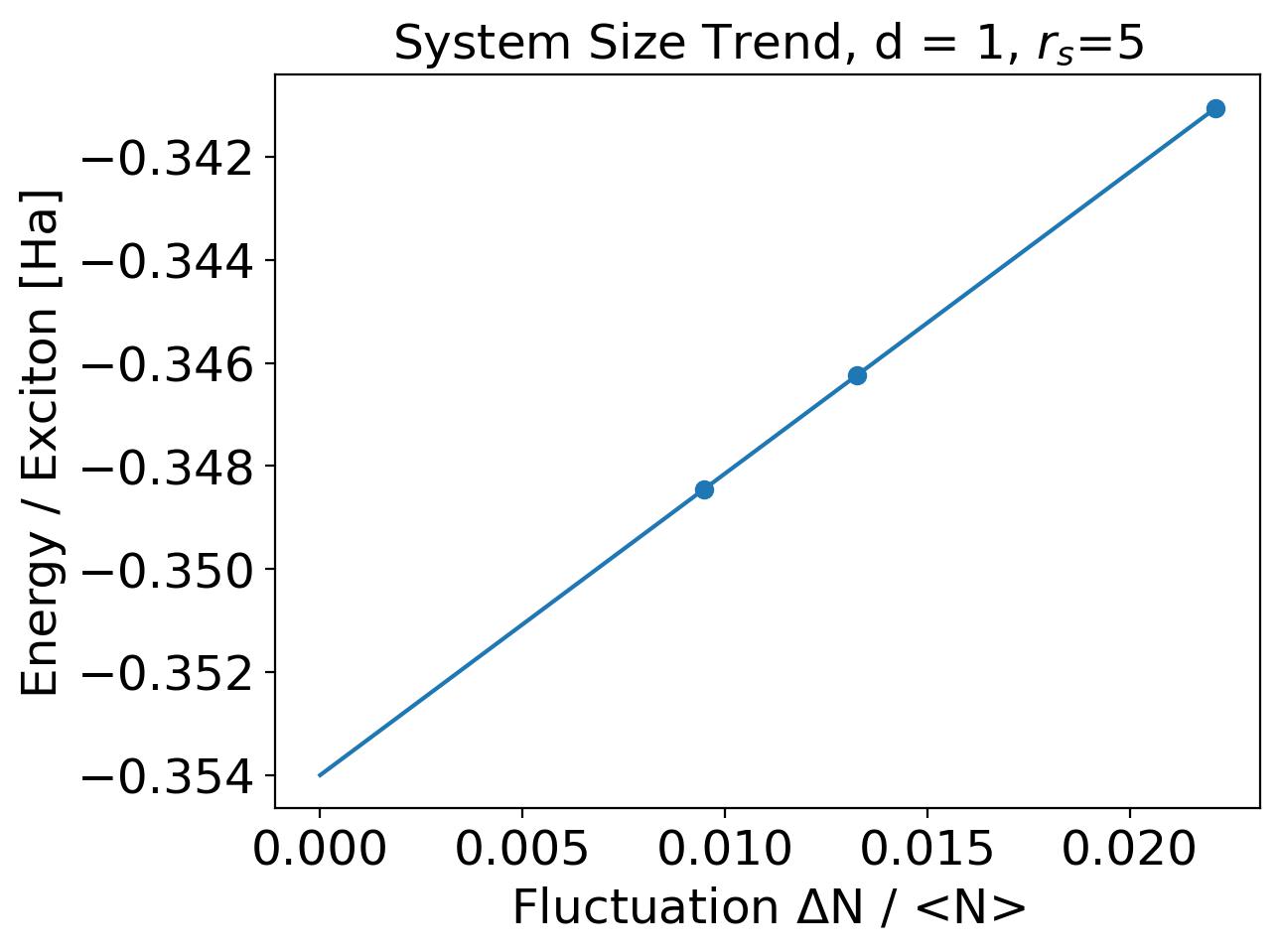}
  \caption{HFB calculations at $d=1, r_s = 5$ and $L = 700, 500, 300$ showing the clear linear trend between the energy per exciton and the relative number fluctuation.}
  \label{fig:trend d1}
\end{figure} 

\section{Neural Network Architecture and Optimization}
We use sin and cos embedding for the particle positions, where each two dimensional position vector is applied with sin and cos function component wise. The parameters in the neural network are initialized using LecunNormal~\cite{klambauer2017self}. The $\v k$ points in our GemiNet include all the $\v k$ points that are within the second closest distance to the origin. The number of iterations $T$ in this works is set to 1. The optimization implemented in JAX~\cite{jax2018github} takes 1000 steps with learning rate $lr=0.15 \times min\{d,r_s \}$. 

\end{document}